\documentclass{article}

\usepackage[
  backend=biber,
  style=numeric-comp,   
  sorting=none,         
  giveninits=true,      
  maxbibnames=99,        
  doi=false,isbn=false,url=false
]{biblatex}
\DeclareFieldFormat[article,inproceedings,incollection]{title}{#1}
\DeclareFieldFormat{journaltitle}{\mkbibemph{#1}}
\DeclareFieldFormat{booktitle}{\mkbibemph{#1}}
\DeclareFieldFormat[article]{volume}{\mkbibbold{#1}}

\usepackage{kotex}
\usepackage[normalem]{ulem} 

\usepackage[nonatbib,dblblindworkshop, final]{neurips_2025}

\workshoptitle{AI for Music}

\usepackage[utf8]{inputenc} 
\usepackage[T1]{fontenc}    
\usepackage{hyperref}       
\usepackage{url}            
\usepackage{booktabs}       
\usepackage{amsfonts}       
\usepackage{nicefrac}       
\usepackage{microtype}      
\usepackage{xcolor}         
\usepackage{graphicx}

\title{From Generation to Attribution: Music AI Agent Architectures for the Post-Streaming Era}

\author{
\textbf{Wonil Kim}$^{1*}$, \textbf{Hyeongseok Wi}$^{1*}$, \textbf{Seungsoon Park}$^{1*}$, \textbf{Taejun Kim}$^{1*}$, \textbf{Sangeun Keum}$^{1*}$, \\[0.2em] 
\textbf{Keunhyoung Kim}$^{1*}$, \textbf{Taewan Kim}$^{1*}$, \textbf{Jongmin Jung}$^{1}$, \textbf{Taehyoung Kim}$^{1}$, \textbf{Gaetan Guerrero}$^{1}$, \\[0.2em]
\textbf{Mael Le Goff}$^{1}$, \textbf{Julie Po}$^{1}$, \textbf{Dongjoo Moon}$^{1}$, \textbf{Juhan Nam}$^{1,2}$, \textbf{Jongpil Lee}$^{1}$\thanks{Equal contribution. Corresponding author: \texttt{ceo@mix.audio}} \\[0.5em]
$^{1}$MixAudio by Neutune,
$^{2}$KAIST \\[0.5em]
}

\begin{document}

\maketitle

\begin{abstract}
Generative AI is reshaping music creation, but its rapid growth exposes structural gaps in attribution, rights management, and economic models. Unlike past media shifts—from live performance to recordings, downloads, and streaming—AI transforms the entire lifecycle of music, collapsing boundaries between creation, distribution, and monetization. However, existing streaming systems, with opaque and concentrated royalty flows, are ill-equipped to handle the scale and complexity of AI-driven production.

We propose a \textit{content-based Music AI Agent} architecture that embeds attribution directly into the creative workflow through block-level retrieval and agentic orchestration. Designed for iterative, session-based interaction, the system organizes music into granular components (\textit{Blocks}) stored in \textit{BlockDB}; each use triggers an \textit{Attribution Layer} event for transparent provenance and real-time settlement.

This framework reframes AI from a generative tool into infrastructure for a \textit{Fair AI Media Platform}. By enabling fine-grained attribution, equitable compensation, and participatory engagement, it points toward a \textit{post-streaming paradigm} where music functions not as a static catalog but as a collaborative and adaptive ecosystem. Session example demo: \url{https://aimusicagent.github.io/}.
\end{abstract}


\section{Introduction}
Generative AI is emerging as a transformative force in the music industry. Advances in cross-modal generation, instrumental synthesis, and voice modeling have enabled music to be created instantly, often in the form of single-shot outputs \cite{dhariwal2020jukebox,copet2023simple,liu2024musicldm}. While this single-shot mode demonstrates technical progress, it limits the depth of creative interaction and obscures the role of attribution. Recent commercial systems (e.g., text-to-music platforms) typically deliver end-to-end songs without offering iterative refinement or mechanisms for crediting contributors. This gap highlights a structural misalignment between technological capability and the requirements of sustainable creative ecosystems. To move beyond this, there is a growing need for systems that treat composition not as a one-time act but as an iterative and participatory process.

\begin{figure}[t]
\centering
\includegraphics[width=\linewidth]{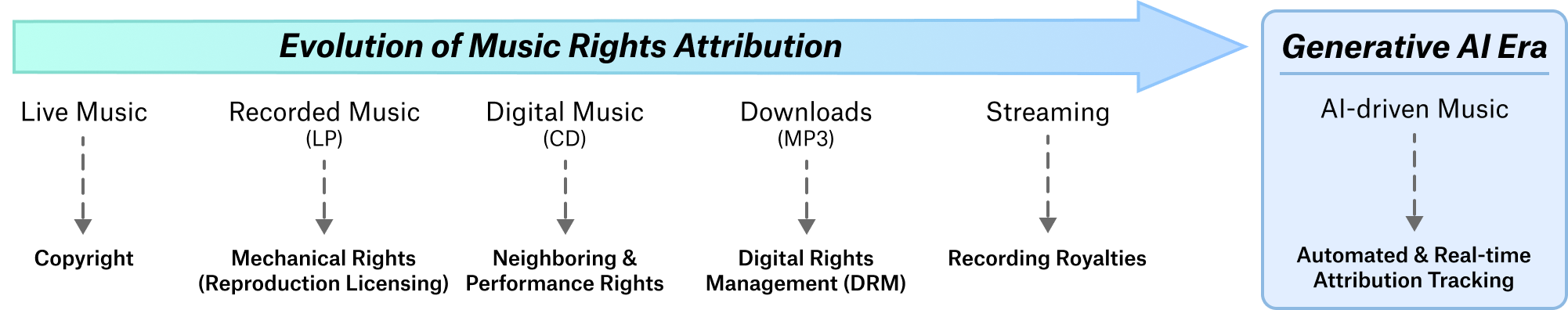}
\caption{Music rights across media shifts into the AI era.}
\label{fig:background}
\end{figure}
\vspace{-0.2em}

As illustrated in Figure~\ref{fig:background}, the history of the music industry has been shaped by successive shifts in media formats, each requiring new legal and economic frameworks. From copyright in live performance to mechanical rights in recorded music, neighboring rights in the CD era, and royalty accounting in streaming, every transition has redefined how rights are attributed and revenues distributed \cite{wipo2021,ifpi2023}. Unlike earlier shifts, which expanded rights around tangible media, generative AI collapses creation, distribution, and monetization into a single process. This convergence creates demands for real-time, fine-grained attribution and accounting frameworks that existing models cannot yet provide \cite{gervais2023ai}.

This gap collides with the structural limitations of today’s streaming economy. Subscription-based royalty models remain opaque and concentrated among a few artists, while metadata fragmentation and fraudulent practices exacerbate the “black box” of unmatched royalties \cite{marshall2015problem,herstand2020streaming,ifpi2023}. As AI dramatically scales both the volume and diversity of music, these inefficiencies risk being magnified: absent attribution-aware infrastructure, independent artists and cultural communities risk seeing their contributions absorbed into opaque accounting with limited recognition or compensation. In other words, AI could enlarge the very inequities it is often expected to solve.

To overcome these limitations, we argue that AI systems must extend beyond generation to encompass attribution, participation, and monetization. Embedding rights tracking not only into the creative process but also into modes of consumption is essential for building sustainable platforms. This trajectory points toward a post-streaming paradigm: an adaptive, participatory, and collaborative ecosystem where music is no longer a static catalog but a living process of creation and engagement.

In this paper, we propose a Music AI Agent architecture that operationalizes this vision. Built on content-based Retrieval-Augmented Generation (RAG), generative models, and agent-based orchestration, the system embeds attribution directly into the creative workflow. By recording \emph{Block-level} usage events—where \emph{Blocks} denote granular musical components such as stems or structural segments—the system directly links creation with transparent provenance. Combined with real-time royalty settlement, this lays the foundation for what can be described as a \textit{Fair AI Media Platform} that aligns generative capabilities with transparent attribution and equitable economic models for the future of music.

\setlength{\parskip}{1pt}

\section{Background}
In parallel with advances in LLMs, RAG, and AI Agents in the language domain, we explore how these paradigms can be extended to media content-based AI Agent systems, with a particular focus on references in the domain of music.

\subsection{Generative Model}
As LLMs serve as foundational models in language, their counterparts in music audio are transformer- \cite{huang2018music,agostinelli2023musiclmgeneratingmusictext, copet2023simple}, diffusion- \cite{schneider2024mousai, liu2024musicldm, 10.5555/3692070.3692575, evans2025stable,evans2024long}, and flow-matching–based \cite{prajwal2024musicflow} generative models, enabling full-song \cite{schneider2024mousai}, stem \cite{parker2024stemgen,rouard2025musicgen,kong2024multi,mariani2024multi}, and singing voice \cite{10.1145/3503161.3547854, 9747664, Liu_Li_Ren_Chen_Zhao_2022, zhuang2024spasvc, zhang2024dsffsvc} generation or conversion \cite{bai2024seed}. Services such as Suno \cite{suno}, Udio \cite{udio}, and Riffusion \cite{riffusion} offer full-song generation but lack transparency in training data and revenue sharing \cite{sunosue}. In response, ElevenLabs \cite{elevenlabs} has partnered with independent music distributors like Kobalt \cite{kobalt} and Merlin \cite{merlin}, and Beatoven.ai \cite{beatovenai} with providers such as Musical AI, though current revenue-sharing models remain rudimentary (e.g., simple 1/N splits) \cite{musicalai}. More recently, Splice has introduced sample generation and transformation tools \cite{splice}, and Moises supports context-aware stem generation—together \cite{moises} signaling both the diversification of AI music services and the need for richer, more interactive user experiences.





\subsection{Content-based RAG}
Just as LLMs have evolved into RAG systems with Memory DBs to support source attribution and richer user experiences \cite{lewis2020rag}, the analogue for media content lies in generative models grounded in a database of component blocks of media \cite{mao2025multirag,shirag2024visrag,zhu2025crossmodalrag,huang2025motionrag}. In these multimodal RAG systems, retrieval is not limited to text but extends across heterogeneous modalities such as visual content and motion data, where textual queries can be jointly matched with image or motion embeddings to enhance specificity of the media content. In music, this translates to a music sample database from which elements are retrieved via text queries and then recombined to generate new content. Such an approach enables explicit reference tracking of samples, facilitating attribution and revenue sharing. Structurally, the mapping aligns with RAG: indexing and parsing correspond to Music Information Retrieval (MIR) \cite{downie2003mir}, while retrieval, language–audio models \cite{primus2025tacos,wang2025u,wijngaard2025audsemthinker,alshammari2025unifying}, and generative models form the pipeline \cite{mariani2024multi,mariani2023multi,xu2024multi,karchkhadze2025simultaneous,karchkhadze2024multi}. 




\subsection{AI Agent Architecture containing Media Content}
Extending content-based RAG with an agent-based orchestration layer and toolsets yields a highly organic user experience in music creation and transformation. Early explorations such as Music AI Agents \cite{yu2023musicagent} integrated diverse MIR techniques into toolsets, enabling users not only to generate music but also to isolate and manipulate its components. However, these systems remain algorithmically toolset-driven \cite{doh2025talkplay}. If evolved into agent architectures backed by media-content memory databases—where samples or artist tracks are decomposed into granular musical elements—such systems could further support attribution and enable revenue-sharing mechanisms proportional to original source contributions.




\setlength{\parskip}{1pt}

\section{Content-based Music AI Agent for Integrated Attribution}

\begin{figure}[t]
    \centering
    \includegraphics[width=\linewidth]{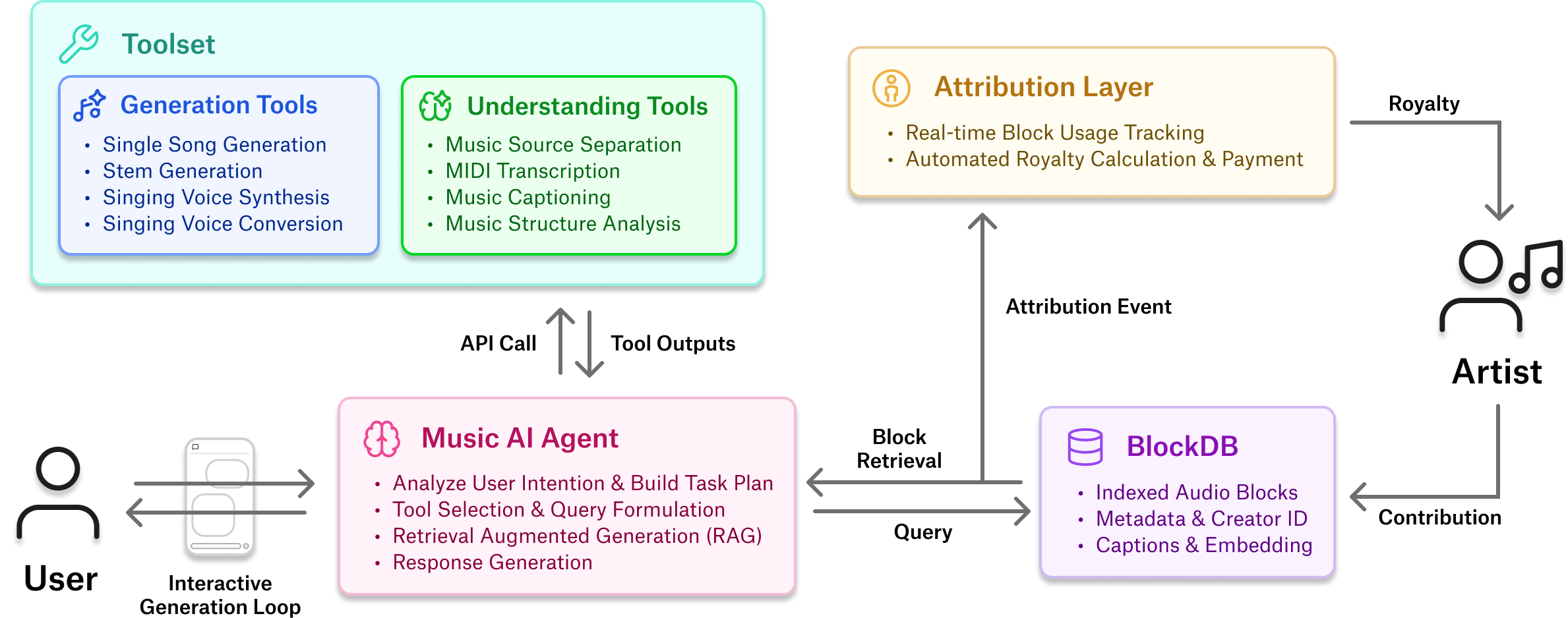}
    \caption{The architecture of the proposed Music AI agent designed to ensure fair artist compensation through an automated attribution and royalty system.}
    \label{fig:system_architecture}
\end{figure}
\vspace{-0.2em}

To address the core challenge of attribution, we introduce an architecture that operationalizes the paradigms of content-based RAG and AI agents. Our system orchestrates a suite of generative tools grounded in a database of musical components, thereby making attribution an intrinsic part of the creative act rather than a post-hoc process. As detailed in Figure \ref{fig:system_architecture}, this architecture is composed of three pillars: a \textbf{BlockDB} of artist-contributed media, an interactive \textbf{Music AI Agent} with a versatile \textbf{toolset}, and a real-time \textbf{Attribution Layer}.


\paragraph{Creator Contribution Workflow}
The process begins with artists submitting original music to the platform, where it is analyzed, separated into \textit{Blocks}, and indexed in BlockDB with metadata including its musical properties, descriptive embeddings, and a record of its original creator. Blocks are created either by decomposing artist tracks—supporting attribution and distribution—or by incorporating stems and samples from producers, which complement the database to address the existing copyright frameworks not covered by decomposition process alone. This process builds the library of attributable musical assets that powers the ecosystem.


\paragraph{Interactive Generation and Real-time Attribution}
The creation process unfolds as a \textbf{session-based, interactive loop}, where users collaborate with the Music AI Agent to build a track layer by layer. In each turn of this creative dialogue, the agent executes a RAG workflow by retrieving relevant Blocks from BlockDB in response to user prompts. These Blocks then serve as context or creative conditions for the diverse generative tools in the agent's toolset. Crucially, each time a Block is used to generate a new musical layer, it triggers a real-time \textbf{attribution event}, linking the new output to its original creators. The Attribution Layer aggregates these events throughout the session and allocates royalties to the original artists under transparent rules, completing the cycle from creation to compensation. By embedding attribution within this iterative RAG pipeline, the architecture ensures that compensation reflects a work's actual granular use, rather than approximations based on market share. This design provides the foundation for a more equitable AI-driven music economy, which is explored further in the following section.

\section{Discussion and Conclusion}

\begin{figure}[t]
  \centering
  \includegraphics[width=\linewidth]{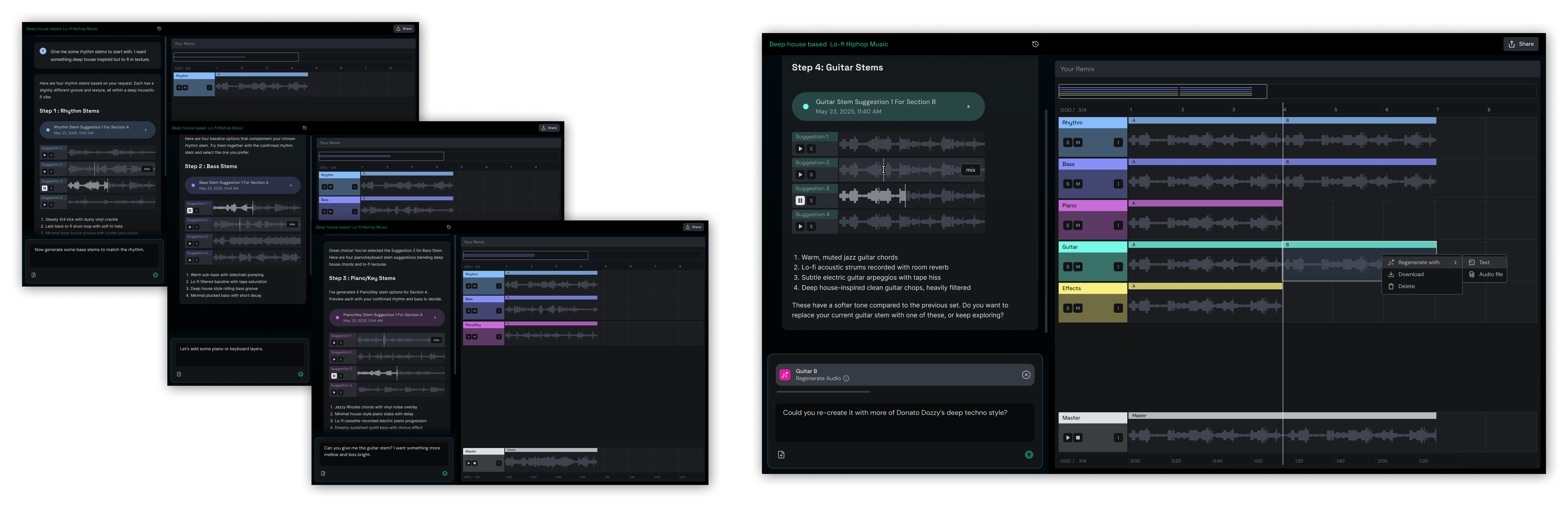}
  \caption{The user interface for a Music AI Agent, showing how a song is built through session-based, interactive layering of musical stems (left), and how a user can then select a specific track to request further modifications (right).}
  \label{fig:agent_ui}
\end{figure}
\vspace{-0.2em}

The introduction of attribution-centered Music AI Agents represents more than a technical innovation. It provides a structural alternative to the existing streaming paradigm and signals the emergence of next-generation AI-driven media platforms.

\paragraph{Transformation of Creative Practice}

If such an agentic workflow is structured as iterative, session-based stem generation as illustrated in Figure \ref{fig:agent_ui}, music creation is reframed as a participatory process that lowers barriers to entry and diversifies engagement. Novices can explore through conversational interaction (e.g., \textbf{vibe producing} or \textbf{remixing}), while professionals can pursue more refined creations with precise control. In this way, music is redefined not as a static product but as a living, adaptive medium of expression \cite{agostinelli2023musiclmgeneratingmusictext,schneider2024mousai}.

\paragraph{AI Attribution}

By embedding attribution directly into the creative process, the system creates verifiable and auditable records at the level of components. This replaces the aggregate and opaque accounting of traditional streaming with transparent, traceable flows of credit and compensation. Every creative action becomes a documented contribution, strengthening accountability among artists, rights holders, and users, and positioning AI as a legally reliable partner in the creative cycle \cite{unesco2023ai}.

\paragraph{Economic Models and Fan Participation}
Granular contribution records enable economic models that move beyond blanket or pool-based subscription royalties. Instead, a usage-based micropayment mechanism could be a solution to current streaming fraud by directly tying rewards to creative contributions. Furthermore, fans evolve from passive listeners to active stakeholders who remix, co-create, and directly support artists. This dynamic expands the scope of the “superfan economy” into a broader ecosystem of participatory patronage \cite{midia2021superfans}.

\paragraph{Infrastructure and Copyright Integration}

Event logging through the Attribution Layer establishes infrastructure compatible with global copyright registries such as ISRC, ISWC, and IPI, positioning Music AI Agents as complementary extensions of existing rights management systems that connect creation, AI generation, and distribution. By managing the music lifecycle through transparent contribution records and automated settlements, Music AI Agents help form the foundation of a Fair AI Media Platform and support a more sustainable, interoperable, and collaborative music ecosystem \cite{mckinsey2023genai}.

\paragraph{Post-Streaming Outlook}

Collectively, these mechanisms signal a shift from centralized streaming platforms toward agentic AI ecosystems where creation, consumption, and settlement converge in real time. In this post-streaming paradigm, AI evolves from a generative tool into a trusted mediator that governs attribution, evaluation, and economic interaction under transparent rules \cite{hesmondhalgh2023music,meier2023poststreaming}.

\section*{Acknowledgements}

This work was made possible through the support of the following institutions and collaborators. \\[0.4em]
\textbf{Institutions:} Korea Creative Content Agency (KOCCA), National Gugak Center, National Information Society Agency (NIA), Korea Culture Information Service Agency (KCISA), Artificial Intelligence Industry Cluster Agency (AICA), and Electronics and Telecommunications Research Institute (ETRI). \\[0.4em]
\textbf{Team:} Theerasak Charoenchob, Teeratep Weerapang, Azamat Atabayev, Abboskhon Sobirov, Nursultan Soodonbekov, Hyeyeon Park \\[0.4em]
\textbf{Advisory:} Yoon Kim, Woojin Cha \\[0.3em]
\textbf{Lead Advisor:} Soo-man Lee \\[0.3em]



{
\small
\printbibliography
}


\appendix

\section{Music AI Agent Pipeline: Architecture and Workflow}

\subsection{System Overview}

This system is an agent-based platform engineered for interactive music generation and understanding, operating on a human-in-the-loop paradigm. The system's architecture is founded upon the synergistic integration of four core pillars: (1) a structured, modular music database, termed \textbf{BlockDB}, which serves as the foundational layer for content retrieval; (2) a versatile suite of \textbf{generative and analytical music tools} for processing and creating a wide range of musical information; (3) a sophisticated \textbf{session-based multi-agent system} that orchestrates the creative workflow by interpreting user intent and managing task execution; and (4) a real-time, automated \textbf{Attribution Layer} that ensures transparent and equitable royalty distribution.

Our architecture prioritizes iterative collaboration, granular control, and transparent attribution. Unlike monolithic models that generate entire tracks from a single prompt, our agent adopts a modular, session-based approach. This method treats music as a structured assembly of components, mirroring how human composers often layer elements like rhythm and melody sequentially. This modularity not only enables a more interactive user experience but also makes complex tasks more controllable by breaking them into smaller, constrained problems, such as generating a single instrumental part. Critically, this session-based paradigm is the foundation for the system's robust attribution mechanism.

\subsection{BlockDB: A Granular Music Component Database}

\subsubsection{The \textit{Block} as a Fundamental Unit}

The atomic unit of musical information within the ecosystem is the "Block." Each Block represents a discrete audio segment of a single instrumental or vocal stem, situated within a specific structural section of a composition. To formalize this concept and enable systematic organization and retrieval, Blocks are categorized along a two-dimensional ontological framework:

\begin{itemize}
\item \textbf{Timbral axis:} This axis classifies Blocks based on their timbral characteristics and instrumental role. For example, a Block could be a distinct instrumental stem such as a percussive drum loop, a bassline played by a bass guitar, a chord progression from a piano, or a melodic synth lead.
\item \textbf{Temporal axis:} This axis provides temporal context by classifying Blocks according to their typical position within a song structure, such as an intro, a verse, or a chorus.
\end{itemize}

This dual-axis categorization imposes a formal structure on musical ideas, transforming a potentially disorganized collection of audio files into a structured repository. This structure functions as a compositional grammar, allowing the AI to reason about abstract musical concepts such as "a high-frequency melodic element for a chorus" as a distinct, queryable object. This moves beyond simple timbre matching to a system that understands compositional context, which is a significant step towards more musically aware AI.

\subsubsection{Enabling Retrieval-Augmented Generation (RAG) and Attribution}

Every Block stored in the database is an atomic entity containing not only the audio data but also a rich set of metadata. This metadata comprises several key components: (1) musical properties, including the Block's timbral and temporal type, BPM, key, and bar length; (2) descriptive text captions; and (3) a suite of embeddings for various modalities. Crucially, the creator of each Block is meticulously recorded. This structured approach facilitates sophisticated \textbf{Retrieval-Augmented Generation (RAG)} and enables robust \textbf{attribution tracking} throughout the creative process. When a Block is retrieved, its information can be used in diverse ways; for instance, its audio can serve as a direct condition for a generation task, or it can be transcribed into a symbolic format like MIDI to provide a melodic condition. In all cases, the retrieval triggers attribution to the original creator.

\subsubsection{Artist Submissions and Content Ingestion}
BlockDB is designed as an open ecosystem that grows through artist contributions, which are ingested via two primary pathways. First, artists can submit finished tracks, which our system then processes through a decomposition pipeline. This pipeline analyzes and decomposes the audio both timbrally into its constituent stems (e.g., vocals, drums, bass) and temporally into structural sections (e.g., verse, chorus). Each resulting component is indexed as a Block with a clear link to the original work, supporting direct attribution. Second, artists and producers can contribute standalone stems and samples directly. This pathway is crucial for incorporating pre-cleared musical components and addressing complex copyright frameworks. All submissions, regardless of type, undergo a systematic ingestion process to be seamlessly integrated as new Blocks, creating a legally robust and creatively diverse library for the platform.

\subsection{A Unified Framework for Generative and Analytical Tools}

\subsubsection{A Diverse and Extensible Toolset}

The agent's core capability is its ability to orchestrate a diverse collection of specialized AI models and conventional signal processing or rule-based tools to accomplish complex user requests. Instead of relying on a single monolithic model, the system leverages a toolset where each tool is optimized for a specific music-related task. This modular toolkit can include, but is not limited to:

\begin{itemize}
    \item \textbf{Generation Tools:} A wide array of models for tasks such as \texttt{single-song-generation} from multi-modal inputs (text, symbolic, audio), \texttt{stem-generation} which creates a new instrumental track from context and prompt audio, \texttt{lyric-to-melody}, \texttt{singing-voice-synthesis} for creating vocals from lyrics and scores, and \texttt{singing-voice-conversion} for altering the timbre of an existing vocal track.
    \item \textbf{Understanding Tools:} Tools designed to interpret and analyze musical content, including models for tasks like \texttt{music-source-separation}, \texttt{MIDI-transcription}, \texttt{music-classification}, \texttt{music-captioning}, and \texttt{music-structure-analysis}.
\end{itemize}

This approach allows the agent to be highly extensible and adaptable. New tools can be integrated into the framework to expand its capabilities without requiring changes to the core architecture.

\subsubsection{Generalized Conditional Framework}

The power of the agent lies in its ability to dynamically chain these tools, using the output of one as the input for another. This process is guided by a generalized, multi-modal conditioning framework. The agent can process and generate various forms of musical information, including:

\begin{itemize}
    \item \textbf{Text:} Descriptive prompts, lyrics, genre tags, or emotional descriptors.
    \item \textbf{Symbolic Music:} MIDI or other sheet music formats representing melody, harmony, and rhythm.
    \item \textbf{Audio:} Raw waveforms containing timbral, stylistic, and performance characteristics.
\end{itemize}

This framework allows for complex and creative workflows. For example, the agent can perform \texttt{stem-generation} by using an existing mix Blocks as context audio and a short instrumental clip Block as prompt audio to add a new, coherent layer to the track. In another common workflow, a user could provide lyrics (text), which the agent uses with a \texttt{lyric-to-melody} tool to generate a MIDI file (symbolic). This symbolic data, along with the lyrics, can then be fed into a \texttt{singing-voice-synthesis} model to produce a complete vocal track. This multi-modal control over different musical aspects offers significant creative flexibility.

\begin{figure}[ht]
  \centering
  \includegraphics[width=\linewidth]{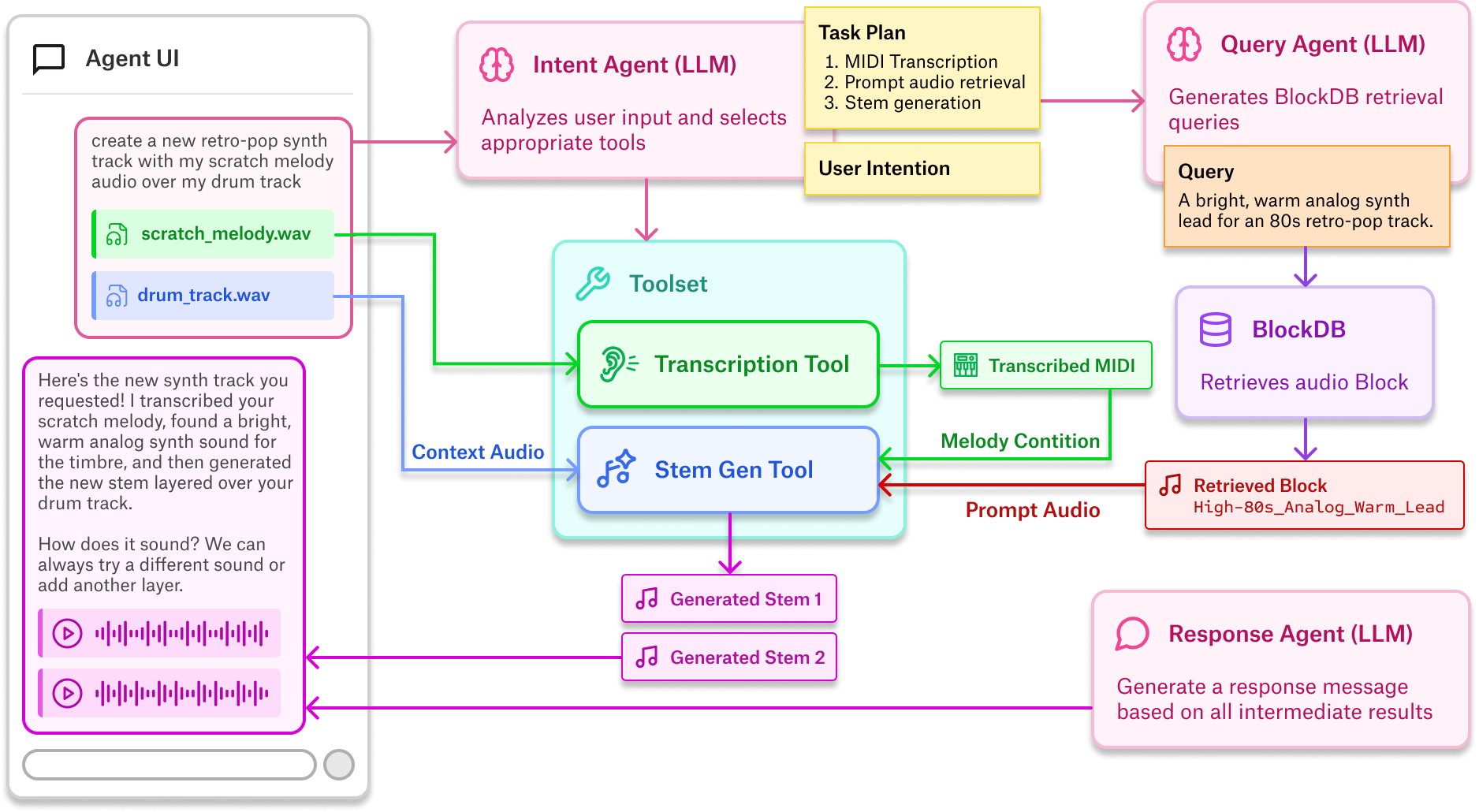}
  \caption{This diagram shows a multi-agent AI's workflow for creating music, where it analyzes a user's prompt and audio files, retrieves a relevant sound Block from BlockDB, and uses various tools to generate new musical stems as a final output.}
  \label{fig:agent-pipeline}
\end{figure}

\subsection{The Music AI Agent: An Interactive and Sequential Workflow}

As depicted in Figure~\ref{fig:agent-pipeline}, the system employs a sequential, interactive methodology orchestrated by a multi-agent architecture. This architecture acts as the central controller for all user-facing tasks, managing user interaction, resource retrieval, and tool execution within an iterative, human-in-the-loop process.

\paragraph{1. Intent Agent: Devising a Task Plan}
The workflow begins when a user submits a prompt. Intent Agent is the first to process this input, analyzing the user's high-level creative goal to devise a direct task plan. This plan is a dependency graph that specifies which tools from the toolset to use and in what sequence. The agent determines how data should flow between these tools to achieve the desired outcome. For example, for a complex request like, "create a new retro-pop synth track with my scratch melody audio over my drum track," the Intent Agent constructs a multi-tool plan:
\begin{enumerate}
    \item First, execute a \texttt{transcription} tool to convert the user's scratch melody audio into a symbolic MIDI file.
    \item Second, retrieve a suitable synth sound from BlockDB to serve as the prompt audio for the timbre.
    \item Third, execute a \texttt{stem-generation} tool, feeding it three distinct inputs: the transcribed MIDI from the first step as the melodic guide, the user's uploaded drum track as the context audio, and the retrieved synth sound from the second step as the prompt audio.
\end{enumerate}

\paragraph{2. Query Agent: Retrieving Attributable Assets}
Following the execution plan from Intent Agent, Query Agent is responsible for gathering the necessary creative assets from \textbf{BlockDB}. This is the retrieval step in the system's Retrieval-Augmented Generation (RAG) process, where the agent translates abstract requirements from the plan into concrete queries. Continuing the example, to fulfill the plan's need for a "retro-pop synth sound," the Query Agent formulates a query to find a suitable audio Block. The retrieved Block is then designated to serve as the prompt audio for the \texttt{stem-generation} tool. This retrieval is the critical link to attribution; the act of fetching a Block immediately triggers a \textbf{Block usage event} to be sent to the Attribution Layer, ensuring the original creator is logged for compensation.

\paragraph{3. Task Execution and Session-based Iterative Loop}
With the execution plan set and all necessary conditional information gathered (from the user or retrieved by the Query Agent), the system executes the specified tools in sequence. The output of one tool, such as the MIDI from the \texttt{transcription} tool, seamlessly becomes an input for the next. The final output, such as newly generated stems, is packaged with a natural language \textbf{response} explaining the process and sent back to the user. This completes one cycle of the interactive loop, allowing the user to provide further prompts to refine the output, add new musical layers, or initiate a different task, creating a collaborative, session-based, human-in-the-loop workflow.

\begin{figure}[ht]
  \centering
  \includegraphics[width=\linewidth]{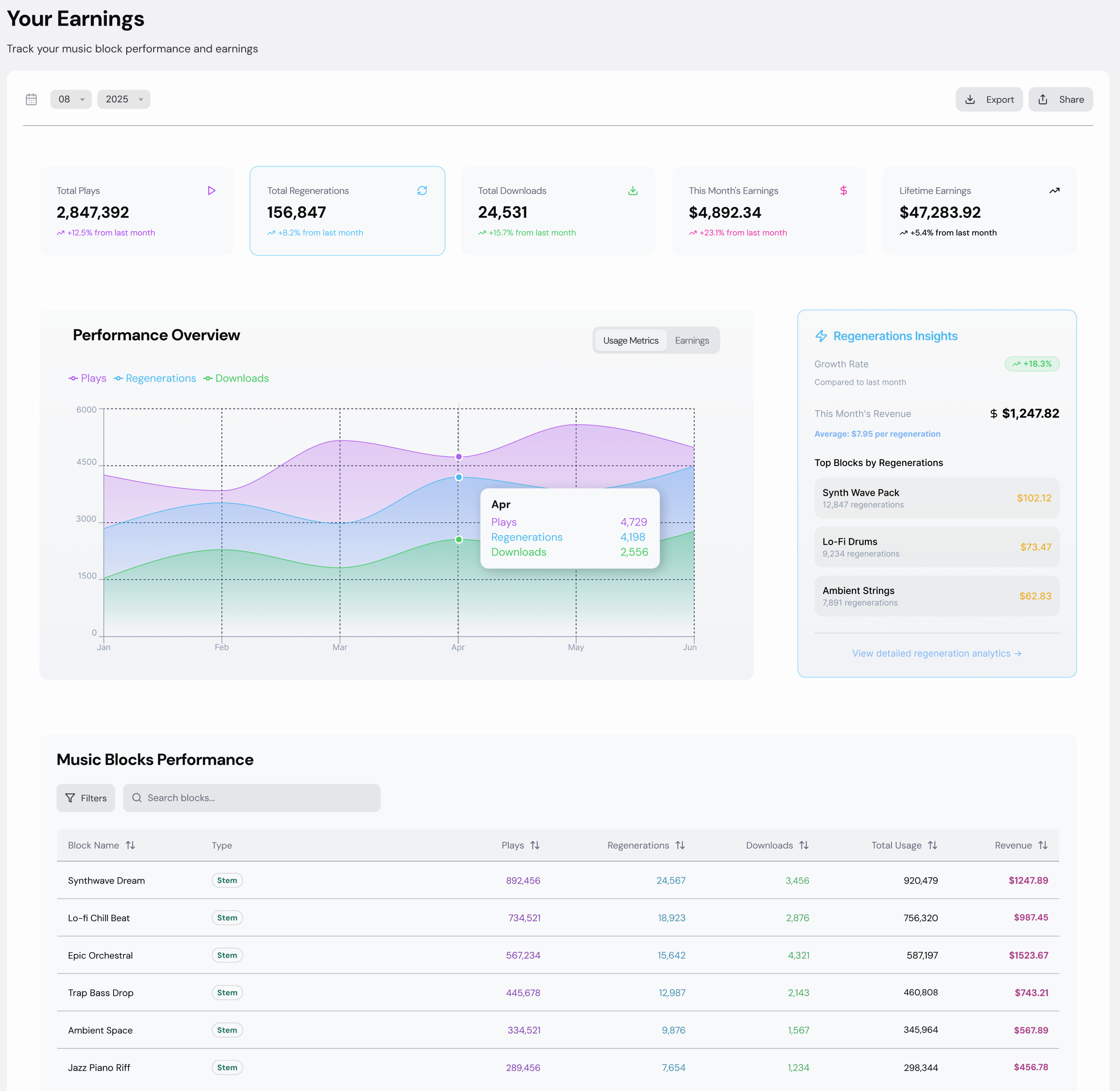}
  \caption{The user interface of dashboard for artists. The dashboard provides detailed analytics on how their individual music Blocks are being used and the revenue they are generating.}
  \label{fig:attribution}
\end{figure}

\subsection{Attribution Layer}

A core principle of this ecosystem is that attribution is not an afterthought but an integral part of the interactive creation process. The system is designed to track and assign credit in real-time. During the interactive loop, every time a Block is retrieved from BlockDB to serve as a condition for a task—whether as direct audio, a source for transcription, or any other form of creative inspiration—the system immediately logs an attribution event for the original creator of that Block. This ensures that artists are credited for the influence and utility of their contributions at the moment of use, creating a direct incentive for populating BlockDB with high-quality, foundational musical ideas (Figure \ref{fig:attribution}).

\subsection{System Extensibility via Model Context Protocol (MCP)}
The system is designed for interoperability with the broader digital music ecosystem. Through Model Context Protocol (MCP) \cite{anthropic2024mcp}, the platform can interface with external environments such as Digital Audio Workstations (DAWs), enabling a seamless and extended creative workflow.


\end{document}